\documentclass[preprint,11pt]{elsarticle}
\usepackage{geometry}
\usepackage{CJKutf8}
\usepackage{amssymb}
\usepackage{siunitx}
\usepackage{amsmath}
\usepackage{caption}
\usepackage{upgreek}

\begin{document}
\captionsetup[figure]{labelfont={bf},labelformat={default},labelsep=period,name={Fig.}}
\begin{CJK}{UTF8}{gbsn}

\begin{frontmatter}

\title{Density-wave like behavior in a new Kagome material Ce$_{2}$Ru$_{3}$Si}

\author{Jinhua Wang}
\author{Shengtai Fan}
\author{Yiwen Li}
\author{Xiyu Zhu\corref{cor1}}
\ead{zhuxiyu@nju.edu.cn} 
\author{Hai-Hu Wen\corref{cor2}}
\ead{hhwen@nju.edu.cn}
\cortext[cor1,cor2]{Corresponding author}

\affiliation{organization={National Laboratory of Solid State Microstructures
		and Department of Physics, Collaborative Innovation Center of Advanced Microstructures },
            addressline={Nanjing University}, 
            city={Nanjing},
            postcode={210093}, 
            state={Jiangsu},
            country={People’s Republic of China}}

\begin{abstract}
Kagome materials with inherent geometric frustration can produce many interesting physical properties, such as flat bands, quantum spin liquid, chiral magnetism, superconductivity and density-wave orders. Sometimes, the localized 4$f$ electrons from Ce atoms coupled with other conduction electrons would also give rise to the flat bands near the Fermi level, and results in the formation of heavy fermion. Thus, it is highly probable that kagome material incorporating Ce element will display nontrivial physical properties. In this study, we present a new Kagome material belonging to the trinary Laves phase, Ce$_{2}$Ru$_{3}$Si, in which kagome plane is formed by Ru atoms. Electrical transport and specific heat measurements reveal a density-wave like transition. A Curie-Weiss behavior is observed in low-temperature region. Meanwhile we also find a relatively large specific coefficient $\gamma_{n}(0)$. The calculated Wilson ratio $R_\mathrm{W}\propto{\chi(0)/\gamma_{n}}$ is approximately 3.1, indicating a moderate electron correlation effect. Chemical doping of Ir at the Ru site rapidly suppresses this density-wave like transition, while Mo doping leads to a gradual decrease in transition temperature. Theoretical calculation indicates both the Ce-4$f$ and Ru-4$d$ electronic bands cross the Fermi level, forming a Mexican-hat-shape Fermi surface close to the Fermi energy, potentially accounting for the observed density-wave like transition. Our findings provide an useful platform for investigating how hybridization between 4$f$ and 4$d$ electrons influences the electronic transport, and the relationship between the density-wave transition and kagome structure.
\end{abstract}

\begin{highlights}
\item A new compound with kagome lattice formed by Ru atoms.
\item A density-wave like transition is discovered in this compound by the measurements of resistivity and specific heat.
\item The hybridization between Ce-4$f$ and Ru-4$d$ electrons, and a Mexican-hat-shape band of Ru may potentially contribute to the density-wave like transition.
\end{highlights}


\end{frontmatter}

\section{Introduction}
The kagome lattice consists of interwoven triangles and hexagons, which inherently leads to geometric frustration\cite{1,2}. The band structure of the ideal kagome lattice exhibits the Dirac point, van Hove singularity (vHS), and flat band\cite{3,4,5}, with the latter two having a high density of states and large effective mass which may lead to strong electron correlations. With strong interaction between charged particles, an instability of Fermi surface may be induced, leading to various quantum phenomenon, such as superconductivity, charge density wave (CDW), pair density wave (PDW), and complex magnetism\cite{5.1}. In contrast, the Dirac point with linear dispersion may results in various topological states\cite{6,7,8,9,10}. Thus, the system with a kagome structure provides fertile ground for studying the interplay between magnetism, electron correlation, superconductivity and topology. In recent years, a new prototype structure based on kagome lattice was discovered, namely $A$V$_{3}$Sb$_{5}$ ($A$ = K, Cs, Rb)\cite{11}, leading to extensive research on interplay and competition between various phases such as superconductivity, CDW, PDW, nematic and stripe order\cite{12,13,14,15,16,17,18,19}. With replacing V by Cr in this structure, a new kagome material CsCr$_{3}$Sb$_{5}$ was found to exhibit strong electron correlation, noteworthy that a flat band exists near the Fermi level; under high pressure, the density-wave transition in this sample was suppressed and unconventional superconductivity was observed\cite{20}. Additionally, the CDW also has been discovered in antiferromagnetic ordered kagome phase FeGe\cite{20.1}, etc..

Compared to 3$d$ elements like V, Cr, Fe, the element Ru possesses 4$d$ electrons, and exhibits relatively weaker electron correlation but quite often shows magnetic feature, which may lead to other novel phenomena. With Ru atoms forming a kagome net, LaRu$_{3}$Si$_{2}$ possesses the highest superconducting transition temperature ($T_\mathrm{c}$) of 7.8 K among other compounds with the same structure, $RT_{3}$Si$_{2}$ and $RT_{3}$B$_{2}$ ($R$ = rare earth metal, $T$ = transition metal), highlighting the significant role of Ru's 4$d$ electrons in forming the superconductivity\cite{21}. Very recently, charge order above room temperature in La(Ru$_{1-x}$Fe$_{x}$)$_{3}$Si$_{2}$ was also discovered\cite{21.1}. Additionally, superconductor CeRu$_{2}$ with C15 Laves phase, in which superconductivity may strongly intertwined with magnetism, features a Ru-kagome plane oriented perpendicular to the [111] direction\cite{22,22.1}. Despite incorporating rare-earth element Ce into the lattice, contrary to predictions, its 4$f$ electrons are not localized; instead, they hybridize with conduction electrons leading to an intermediate valence state for Ce ions\cite{23}. This hybridization involving 4$f$ electrons can give rise to various correlated electronic phenomena such as heavy fermion, mixed-valence system, and unconventional superconductivity\cite{24,24.1}. 

The Laves phase is characterized by three predominant structural types, the hexagonal C14 type (MgZn$_{2}$-type), the cubic C15 type (MgCu$_{2}$-type), and the hexagonal C36 type (MgNi$_{2}$-type)\cite{25,26}. Traditional Lave phase includes binary compounds with common formula of $AB_{2}$, however if substituting non-kagome-transition-metal element with non-metallic element Si will generate trinary Lave phase, such as Sc$_{2}$Ir$_{4-x}$Si$_{x}$, Mg$_{2}$Ni$_{3}$Si, and superconductor Mg$_{2}$Ir$_{3}$Si\cite{26.1,26.2}. In this study, we present a new Ru-based kagome material with the hexagonal structure: Ce$_{2}$Ru$_{3}$Si. Transport measurements reveal a shoulder-like behavior in the temperature dependent resistivity curve, indicative of a potential density-wave like transition; this finding is corroborated by specific heat measurements. Meanwhile, the magnetization curve basically obeys the Curie-Weiss law, and no obvious anomaly is observed. This discrepancy may be attributed to an enhanced delocalization degree of the 4$f$-electrons due to their hybridization with conduction electrons. Density functional theory (DFT) calculations indicate that both the 4$f$-electron bands from Ce ions and the 4$d$-electron bands from Ru ions cross the Fermi level; notably, the 4$d$ band from Ru forms a Mexican-hat-shaped band near the Fermi level, which gives rise to a vHS.

\section{Experimental method}
The polycrystalline sample Ce$_{2}$Ru$_{3}$Si was synthesized using an arc-melting method with stoichiometric amounts of elements, Ce chips cut from a chunk (purity 99.9$\%$, Cuibolin), Ru powder (purity 99.9$\%$, Alfa Aesar), and Si powder (purity 99.99$\%$, Alfa Aesar). They were weighed and thoroughly ground into powder and pelletized. The pellet was arc melted in multiple rounds with the pellet up and down in each melting to ensure homogeneity. After that, they were wrapped in Ta sheet, sealed in an evacuated quartz tube under vacuum, then annealed at \qty{900}{\degreeCelsius} for more than 30 hours. To guarantee the accuracy of resistivity measurements, the sample was cut into regular bar shape before the transport measurements. Doped samples containing Ir or Mo were prepared by identical procedures.  

The powder X-ray diffraction (XRD) patterns are collected on a Bruker D8 Advanced diffractometer with Cu $K_{\alpha}$ radiation at room temperature. Rietveld refinement fitting of all patterns are performed by using the TOPAS 4.2 software\cite{27,28}. The electrical and thermal transport measurements are conducted on the physical property measurement system (PPMS-9 T, Quantum Design) through typical four-probe method with applied field perpendicular to the ground. The magnetic properties are measured by the superconducting quantum interference device with a vibrating sample magnetometer (SQUID-VSM 7 T, Quantum Design). 

The electronic structure of Ce$_2$Ru$_3$Si was calculated using the plane-wave density functional theory within projector-augmented wave scheme as implemented in Vienna Ab initio Simulation Package (VASP)\cite{29,30,31,32}. The generalized gradient approximation (GGA) in the Perdew-Burke Ernzerhof (PBE) form was used in the self-consistently calculation of the electronic structure\cite{33}. Here, we set a $7\times7\times7$ k-point grid with a cutoff energy of 400 eV and the lattice constants were taken from the experimental structural refinement by TOPAS 4.2 software.

\section{Results and discussion }

\begin{figure*}[htbp]
    \centering
	\includegraphics[width=10cm]{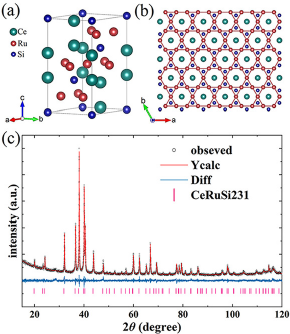}
	\caption{(a) The crystal structure of Ce$_{2}$Ru$_{3}$Si. The green, red and blue spheres represent Ce, Ru and Si atoms, respectively. (b) Top view of the Ru kagome plane. (c) Powder X-ray diffraction patterns (circles) and corresponding Rietveld fitting curve (red solid line) of Ce$_{2}$Ru$_{3}$Si.}
	\label{fig1}
\end{figure*}

Figure. \ref{fig1}(a) illustrates schematic crystal structure of Ce$_{2}$Ru$_{3}$Si, where the green, red and blue spheres represent the Ce, Ru and Si atoms, respectively. It crystallizes in the hexagonal symmetry (space group: R$\bar{3}$m) with Ru kagome plane, as depicted in Fig. \ref{fig1}(b), and Ce/Si rhombohedral plane stacking regularly along the $c$ axis, featuring the same structure with Mg$_{2}$Ni$_{3}$Si. The powder XRD patterns alongside relevant Rietveld refinements are presented in Fig. \ref{fig1}(c). The Rietveld refinements agreement factors $R_{wp}=6.63\%$ and $R_{p}=5.18\%$ are obtained, and samall values indicate that the calculated profile agrees with our experimental date quite well. The lattice parameters are derived with $a=b=5.6125(4)$ \AA, $c=11.4609(5)$ \AA. More detailed results of Rietveld refinements are listed in Table \ref{table1}. All peaks in the XRD patterns are well indexed, mainfesting the good quality and purity of sample.

\begin{table}
	\centering
	\begin{tabular}{lllllll}
		\hline
		\multicolumn{3}{l}{Compound}  & \multicolumn{3}{l}{Ce$_2$Ru$_3$Si}  \\
		\hline
		\multicolumn{3}{l}{space group    }          & \multicolumn{3}{l}{$R\bar{3}m$}   \\
		\multicolumn{3}{l}{$a$ ($\mathring{A}$) }    & \multicolumn{3}{l}{5.6125(4)}  \\
		\multicolumn{3}{l}{$c$ ($\mathring{A}$) }    & \multicolumn{3}{l}{11.4609(5)}  \\
		\multicolumn{3}{l}{$V$ ($\mathring{A}^3$)  }   & \multicolumn{3}{l}{312.6581(1)}  \\
		\multicolumn{3}{l}{$\rho$ (g/cm$^3$)  }   & \multicolumn{3}{l}{9.743}  \\
		\multicolumn{3}{l}{$R_{wp}$ (\%) }           & \multicolumn{3}{l}{6.63} \\
		\multicolumn{3}{l}{$R_{p}$ (\%) }           & \multicolumn{3}{l}{5.18} \\
		\multicolumn{3}{l}{$GOF$}           & \multicolumn{3}{l}{1.25} \\
		\hline
		Atom & Site & $x$ & $y$ & $z$ & Occupancy & B$_{eq}$ \\
		\hline
		Ce  & 6c & 0 & 0 & 0.3686(2) & 1 & 2.079 \\
		Ru & 9d & 0.5 & 0 & 0.5 & 1 & 1.705 \\
		Si & 3a & 0 & 0 & 0 & 1 & 0.7756 \\
		\hline
	\end{tabular}
	\caption{Crystallographic data of Ce$_2$Ru$_3$Si at 300 K.}
	\label{table1}
\end{table}

\begin{figure*}[htbp]
	\includegraphics[width=\linewidth]{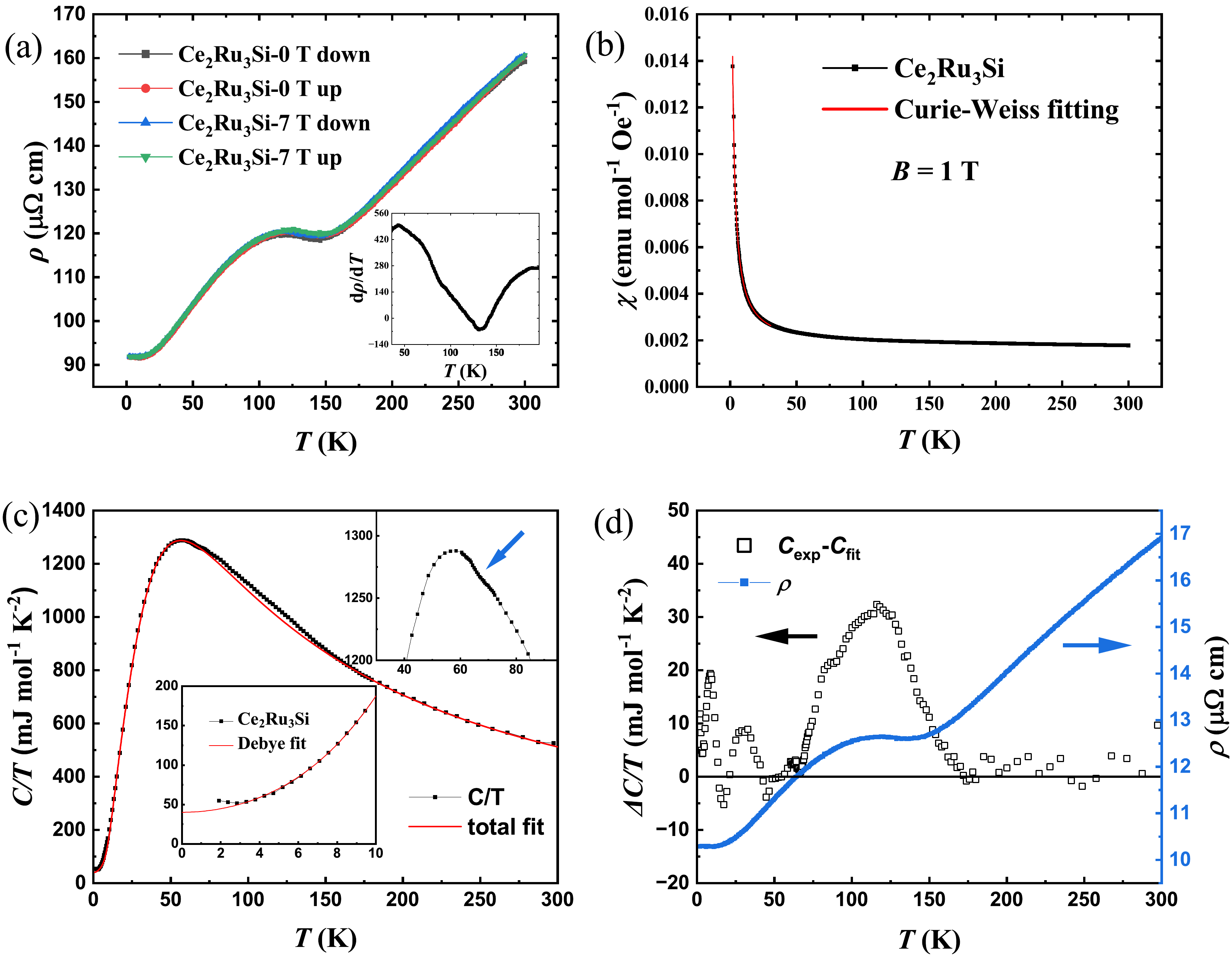}
	\caption{(a) The temperature dependence of resistivity measuring with temperature increasing and decreasing under 0 and 7 T, respectively. The inset shows the differential result of resistivity on warming up at zero field. (b) The temperature dependent magnetization under an applied field of 1 T. Red solid line represents the Curie-Weiss fitting result using the equation of $\chi(T)=\chi(0)+C\/(T+T_{0})$. (c) Specific heat measurements and corresponding fitting curve with a model of $C(T)=a*D(\theta_{D,T})+b*E(\theta_{E_{1}},T)+c*E(\theta_{E_{2}},T)+\gamma_{n}T$. The lower inset exhibits low-temperature specific heat fitting results with a simple Debye model, $C\/T=\gamma+\beta T^{2}+\delta T^{4}$. The upper inset represents beginning of density-wave like transition. (d) The difference between $C_{exp}$ and $C_{fit}$ alongside with resistivity under 0 T.} 
	\label{fig2}
\end{figure*}

Figure \ref{fig2} presents the fundamental physical properties of Ce$_{2}$Ru$_{3}$Si. Fig. \ref{fig2}(a) shows the temperature dependence of resistivity. A broad shoulder-like behavior occurs at around 150 K. When decreasing temperature, the resistivity forms a clear shoulder starting from about 150 K with the maximum at about 125 K, then it decreases continuously again. We use the criterion of the minimum value of the differential resistivity to define the transition temperature, as shown in the inset of Fig. \ref{fig2}(a), and the derived transition temperature is 137 K. Similar behavior was observed previously in Cu-doped TiSe$_{2}$, in which it was ascribed to the charge-density-wave phase transition\cite{34}. Here, the nature of this shoulder is ambiguous, but it is most possibly the result of density-wave transition. Further experiments are required to explore the underlying mechanism. Apart from this anomaly, the sample exhibits metallic property throughout the entire temperature range, and shows negligible magnetoresistance. Contrasting to the electrical transport measurements showing unusual behavior, it seems to be featureless in magnetic measurement besides a Curie-Weiss like paramagnetic feature in whole temperature region. The curve basically conforms to the Curie-Weiss law. From the fitting in low-temperature range, by using the equation of $\chi(T)=\chi(0)+C\/(T+T_{0})$, we can derive the effective moment $\mu_\mathrm{eff}=0.48\mu_\mathrm{B}$, and the Pauli magnetic susceptibility is about $\chi(0)=0.00171$ emu mol$^{-1}$  Oe$^{-1}$. The small value of effective moment suggests delocalization of 4$f$ electrons of Ce. 

Figure \ref{fig2}(c) presents the measured specific heat and its corresponding fitting results. Here, we employ a model given by $C(T)=a*D(\theta_{D,T})+b*E(\theta_{E_{1},T})+c*E(\theta_{E_{2},T})+\gamma_{n}T$ to fit the experimental data\cite{35,36,37}. This model accounts for phonon contributions through an original Debye mode $D(\theta_{D},T)$, two Einstein modes $E(\theta_{E_{1}},T)$ and $E(\theta_{E_{2}},T)$\cite{38}, as well as a linear term representing the electron contribution $\gamma_{n}T$. In the formula, coefficients $a,b$ and $c$ are adjustable parameters of three modes, constrained by the condition $a+b+c=1$. With fixed value of $\gamma_{n}$ and $\theta_{D}$ calculated from low-temperature fitting, we obtain the final fitting curve, as the red solid line shown in Fig. \ref{fig2}(c). A deviation between the experimental data $C_{exp}$ and the fitting curve $C_{fit}$ can be seen from 66 K to 171 K. The beginning of this discrepancy is highlighted in the upper inset of Fig. \ref{fig2}(c). The lower inset of Fig. \ref{fig2}(c) depicts the simple Debye fitting at low-temperature range, using formula $C\/T=\gamma+\beta T^{2}+\delta T^{4}$, whereby the first term comes from electrons, and the latter two terms are contributed by phonons. Ultimately, we get the Sommerfeld constant $\gamma_{n}=40.05$ mJ mol$^{-1}$ K$^{-2}$, and the Debye temperature $\theta_{D}=218.6$ K. The electronic specific coefficient is relatively large. The Wilson ratio is frequently used to evaluate the strength of correlation of electrons in the material, and it can be written as $R_\mathrm{W}=(4\pi^2\times k_{B}^{2})\chi(0)/(3{g\mu_{B}}^{2})\gamma_{n}$, where $k_{B}$ is the Boltzmann constant, $g$ is the Lande factor with numerical value of 2, $\mu_{B}$ is the Bohr magneton\cite{38.1}. Combining the fitting results of magnetization and specific heat, we determine the Wilson ratio which is $R_\mathrm{W}=3.1$, being larger than 2 commonly deemed that the sample possess relatively strong electronic correlation. The difference between these curves ($C_{exp}-C_{fit}$) along with resistivity measurement result at zero magnetic field are presented in Fig. \ref{fig2}(d). The anomaly becomes pronounced as indicated by a peak of $C_{exp}-C_{fit}$, consistent with the transition evident in the resistivity curve, which further confirms the existence of density-wave like transition in the sample. 

\begin{figure*}[htbp]
	\centering
	\includegraphics[width=\textwidth]{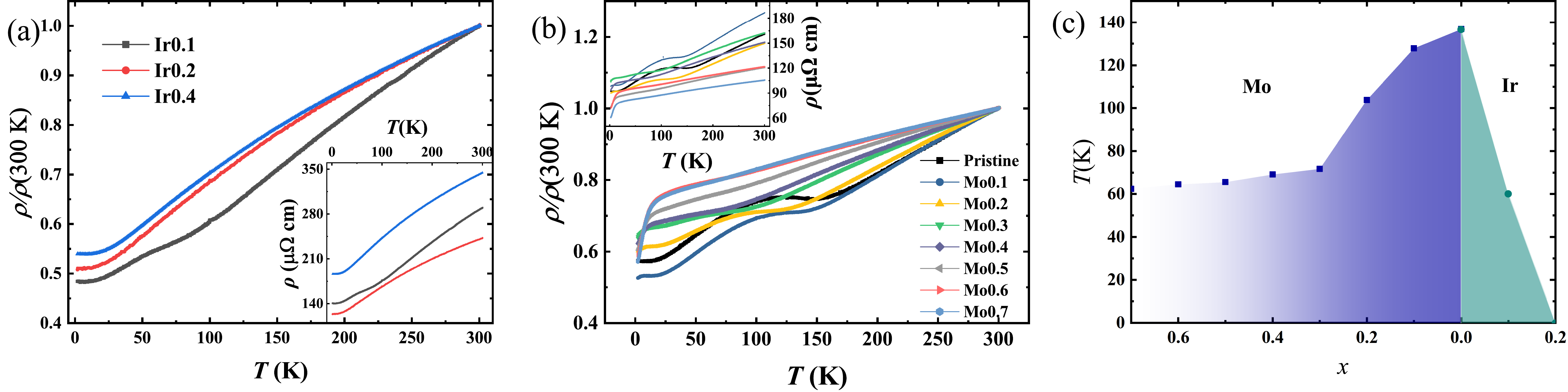}
	\caption{(a) Normalized resistivity of different doped concentrations of Ir. The inset depicts individual resistivity of these sample. (b) Normalized resistivity of various doping levels of Mo. The inset shows corresponding individual resistivity. (c) Phase diagram of transition temperature against doping levels of Ir and Mo.
	} 
	\label{fig3}
\end{figure*}

To investigate this transition further, we conduct chemical doping on Ce and Ru site in order to suppress the transition, and explore potential non-trivial physical properties such as superconductivity. Substitution of La on Ce site (Ce$_{2-x}$La$_{x}$)Ru$_{3}$Si can indeed suppress the transition slightly, with the doping level of $x=0.2$, as shown in Fig. \ref{Sfig1}(a). However, when the concentration of La is increased to 0.4, the crystal structure transforms to cubic C15 type Laves phase, as presented in Fig. \ref{Sfig1}(c), and some peaks of impurity arise. Fig. \ref{fig3}(a) displays transport properties of Ir-doped sample Ce$_{2}$(Ru$_{3-x}$Ir$_{x}$)Si with different doping levels. At an Ir doping concentration of $x=0.1$, there is a notable reduction in transition temperature from 137 K (in pristine sample) to 60 K. Slightly increasing the doping level to $x=0.2$ leads to complete disappearance of the density-wave like transition, but unfortunately no superconductivity was observed. Further increasing the doping level does not alter transport characteristics significantly, and they predominantly exhibit Fermi liquid behavior as shown in individual curve of temperature dependent resistivity, as depicted in the inset of Fig. \ref{fig3}(a). Additionally, we explored Mo substitution for Ru atoms. The resistivity data normalized by the one at 300 K for various doping levels are shown in Fig. \ref{fig3}(b), with corresponding resistivity displayed in its inset. Similar to the doping effect of Ir, Mo substitution also suppresses the transition, but with a much weaker effect compared to Ir. Under about $3\%$ replacement, the small resistivity drop shows up at low-temperature range. The decline of resistivity keeps increasing with incrementally doped Mo, while this bending down of resistivity in Mo-doped samples cannot be attributed to superconductivity, since there is no diamagnetic magnetic signal was observed. We argue that it is rather due to the antiferromagnetism (AFM)-a characteristic signature stemming from the impurity phase like CeRuSi\cite{39}. With doping level of $20\%$, there still exhibits slight density-wave transition, only accompanied with transition temperature down to 60 K. In the sample with a lower doping level of Mo, the amount of this Ce111 phase is below the X-ray error bar of diffraction measurement, yielding no observable impurity peaks in XRD patterns, yet the resistivity still exhibits AFM signature, as presented in Fig. \ref{Sfig2}. Utilizing identical criteria established for defining transition temperature in pristine sample of the density-wave transition, we construct a phase diagram of the transition temperature versus doping levels of Ir and Mo, which is depicted in Fig. \ref{fig3}(c). The transition temperature monotonically decreases with increasing doping level of Ir, while it plateaus under higher Mo substitutions possibly caused by its upper limit of doping.

\begin{figure*}[htbp]
    \centering
	\includegraphics[width=\linewidth]{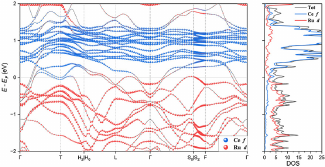}
	\caption{Electronic band structure calculated through the plane-wave density functional theory (DFT), alongside with density of states shown on the right-hand side.}
	\label{fig4}
\end{figure*}

In order to get a comprehensive understanding on the electronic properties of Ce$_{2}$Ru$_{3}$Si with perfect kagome planes of Ru atoms, we try to investigate its electronic structure through the plane-wave density functional theory (DFT) calculation. Considering the presence of heavy atom Ce, it is necessary to include the spin-orbital coupling during the calculation. The band diagram is showcased in Fig. \ref{fig4} alongside density of states (DOS). Although the sample containing element Ce, the flat bands, typically associated with hybridized 4$f$ electrons, remain distanced from the Fermi level, whilst only a small electron pocket exists at T point near the Fermi level. Thus, the 4$f$ electrons only have contribution to the conduction, while no heavy fermion behavior takes place in the sample. Meanwhile, there are few bands originating from Ru 4$d$ electrons crossing the Fermi level, consequently forming both electrons and holes pockets participating overall conductivity, characterizing multiband features. Due to the kagome structure of Ru, some typical band signatures can be observed in the diagram, including a saddle point along H$_{0}$-L direction and a clear Dirac point located at L point proximity to the Fermi level. Interestingly, a Mexican-hat-shape band exists at $\Gamma$ point, giving rise to a vHS in the DOS at the band edge. The Mexican-hat-shape band normally can induce the correlation effect, superconductivity and other phase transition, which has been discovered in magic-angle twisted bilayer graphene, kagome magnet Co$_{3}$Sn$_{2}$S$_{2}$ and topological insulator Sn-doped Bi$_{1.1}$Sb$_{0.9}$Te$_{2}$S\cite{42,41,49,50,51,52,53}. Thus, this material warrants further studies to induce non-trivial phase transitions through moving the vHS closer to the Fermi level.

\section{Conclusions}
We synthesis a new compound Ce$_{2}$Ru$_{3}$Si showcasing kagome structure of Ru. A broad shoulder-like anomaly is observed in electric transport, which may be attributed to a density-wave like transition, further corroborated by specific heat measurement. However, no obvious sign of phase transition occurs in magnetization measurement, ruling out the possibility of spin density wave. This density-wave like transition is very likely related to charge density wave. Based on the fitting results from magnetization and low-temperature specific heat analysis, the obtained Wilson ratio is 3.1, suggesting moderate correlation of electrons in the system. According to the band diagram performed through DFT theory, 4$f$ electrons of Ce does not manage to form flat bands near the Fermi level, on the contrary, only a small electron pocket exists near the Fermi level. The calculation reveals the existence of a Dirac point, saddle point, and notably a Mexican-hat dispersion in the system. Our findings provide a new platform to investigate the interplay between density-wave phase transition and kagome structure, at the same time, the relationship of conducting 4$f$ electrons and electronic correlation.

\section*{Data availability}
Data will be made available on request.

\section*{Acknowledgements}
This work was supported by the National Key R$\&$D Program of China (No. 2022YFA1403201), National Natural Science Foundation of China (Nos. 12061131001, 11927809, 52072170, 52472276), Fundamental Research Funds for the Central Universities (Grant No. 2024300350), and the Strategic Priority Research Program (B) of Chinese Academy of Sciences (No. XDB25000000). 


\clearpage
\begin{center}
	\textbf{\large Supplementary materials for Density wave like behavior in a new Kagome material Ce$_{2}$Ru$_{3}$Si}\\[.15cm]
\end{center}
\renewcommand{\thefigure}{S\arabic{figure}}
\setcounter{figure}{0}

\begin{figure}[h]
	\centering
	\includegraphics[width=14cm]{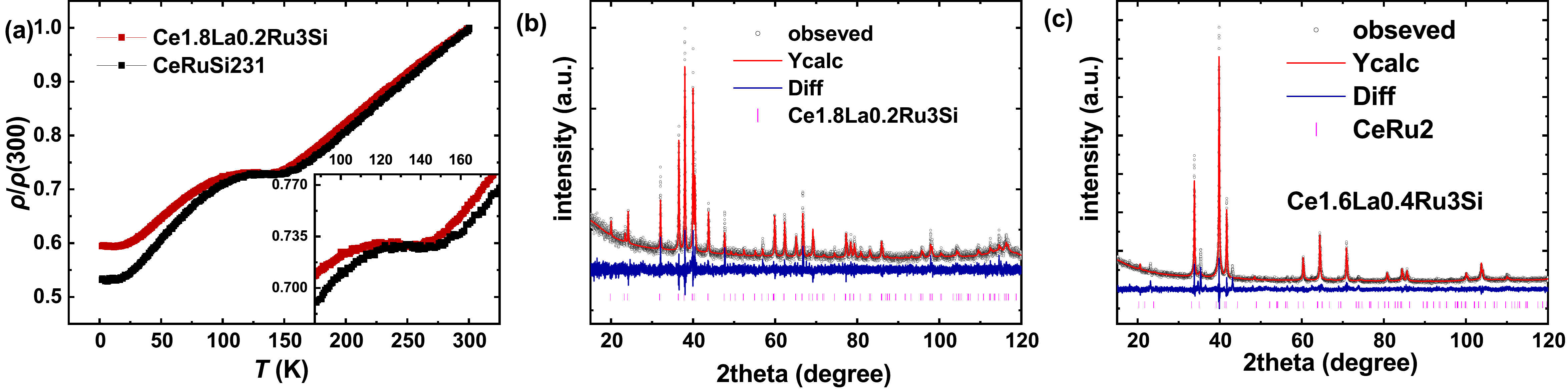}
	\caption[1]{(a) Normalized resistivity of La-doped and pristine Ce$_{2}$Ru$_{3}$Si. Inset shows the enlarge view near the transition. XRD patterns and Rietveld refinements with doped level of (b) 0.2 and (c) 0.4. } 
	\label{Sfig1}
\end{figure}

\begin{figure*}[ht]
	\centering
	\includegraphics[width=14cm]{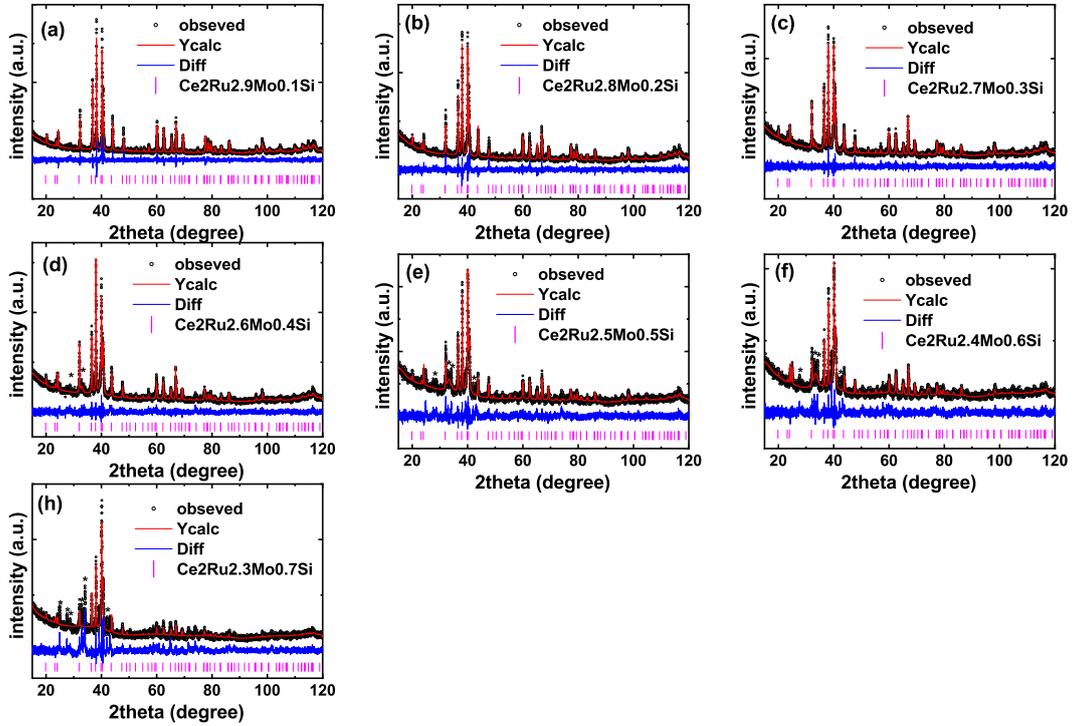}
	\caption[2]{(a)-(h) XRD patterns and Rietveld refinements of various doped level of Mo. Peaks marked by stars belong to the phase of CeRuSi.} 
	\label{Sfig2}
\end{figure*}
\end{CJK}
\end{document}